# Policy Specification in Role based Access Control on Clouds

Gitanjali
Department of CSE
GNDEC, Ludhiana

Sukhjit Singh Sehra
Department of CSE
GNDEC, Ludhiana

Jaiteg Singh
Department of CSE
Chitkara Institute,Patiala

## ABSTRACT

Cloud Computing is a set of IT Services that are provided to a customer over a network and these services are delivered by third party provider who owns the infrastructure and reduce the burden at user's end. Nowadays researchers devoted their work access control method to enhance the security on Cloud. RBAC is attractive access model because the number of roles is significantly less hence users can be easily classified according to their roles. The Role-based Access Control (RBAC) model provides efficient way to manage access to information while reducing the cost of security administration and complexity in large networked applications. This paper specify various policies in RBAC on clouds such as migration policy which helps the user to migrate the database schema and roles easily to the Cloud using XML with more security. Restriction policy provide the security enhancement in Role Based Access Model by restricting the number of transaction per user and if the number of transactions will increase the admin will come to know through its monitoring system that unauthorized access has been made and it would be easier to take action against such happening. This paper proposes backup and restoration policy in Role Based Access Model in which if the main cloud is crashed or not working properly then the backup and restoration facility will be available to avoid the lost of important data. In this case chances of loss of data are very less so enhance more security on Cloud Computing.

## 1. INTRODUCTION

Cloud computing is the boom in the field of the development and application modification in this modern world. In the previous age of development user use to create applications on the local server and use to keep them on the local server. If the local server crashes, the entire system and the applications crashes automatically. It was getting into a huge problem all over the world. To overcome the problem of server crashing, cloud computing was brought into action. Brand companies like Google, Microsoft and facebook have their own clouds where data is available in bulk. As Cloud computing is becoming more popular many business organizations are migrating to Cloud. It requires minimum investment for any business to run on Clouds. Cloud Computing is a technology which provides software, data access, and storage services and that even does not require user to know the physical location and configuration of the system that delivers the services. To provide the security, fast access and privacy are the big challenges in Cloud Computing. In the cloud, due to multi-tenancy architecture, data from multiple clients are stored and managed by the same software[1] . When the software makes mistake, clients may access private data of other clients. Furthermore, data stored in a cloud may be available to cloud administrators and they may modify data for their own benefits. The multiuser environment has increased security risk due to the sharing of software and data schemas by multiple users. This is the responsibility of the cloud providers to ensure that one customer cannot break into another customer's data and applications. Access Control methods in Cloud basically allows access only to the authorized person. It is mechanism that provide the security so that user cannot access the resources for which user is not authorized. Various access control models are in cloud computing are Mandatory Access Control (MAC), Discretionary Access Control (DAC) and Role Based Access Control (RBAC) [2]. Access control models is a means by which ability is explicitly enabled or restricted in some way. Computer based access control models can prescribe who have access to a specific system resource, but also the type of access that is permitted.

## 2. ROLE-BASED ACCESS CONTROL

Role Based Access Control (RBAC) has two phases to assign a privilege to a user [4]. As shown in figure 1, In first phase, one or more roles are assigned to the users. In second phase, the roles are checked against the requested policies or operations. In RBAC, permissions are not associated with user and it is associated with the roles. Roles may have a hierarchical structure, reflecting the organization lines of responsibilty and authority.

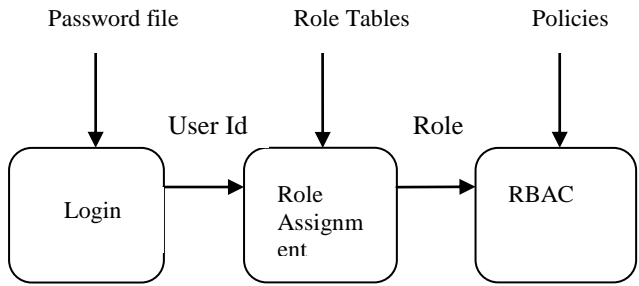

**Figure 1: Two Phase Role Based Access Control Model**

In a RBAC model, all grant authorizations deal with roles. Users are then made members of roles and acquiring the roles' authorizations. User access to resources is controlled by roles; each user is authorized to play certain roles and, based on his own role he can perform accesses to the resources and operate them correspondingly. As a role organizes a set of related authorizations together, it can simplify the authorization management. Whenever a user needs a certain type of authority to perform an activity, user has to be granted the authority of a proper role, rather than directly assigned the specific authorizations. Furthermore, with Role-Based Access Control, decisions are based on the concept of user groups in access control. Roles are closely related to individual users have in an organization. Role based Access Control principles include: separation of duties, data abstraction and least privilege.

## 3. POLICY SPECIFICATION

Policies are derived from business goals and service level agreements in enterprises, which are "rules governing the choices in behaviour of a system [15]. Policy is the plan and course of action that should be following and intended to determine decision, actions and matters. Policies include Access Policy, Authorization and obligation , Migration Policy, Restriction Policy, Backup and restore policy. There is improvement in RBAC by proposing three more policies as an extension of policy specification in the access control model.

### 3.1 Access Policy

Access Policy defines what type of access has been given to the particular user. Read, Write, delete are the access given to the user according to their role. Like admin has all access rights but employee may have an access to read but not to write any information without the admin permission.

### 3.2 Authorization and Obligation Policy

Authorization policy helps to provide the secure architecture. Privacy and Security are the two big issues in any environment . Authorization policy check whether the user is authorized to access the resource or not. If person can access the resources then it give the positive result else give negative result. This help to decide whether the access are allowed or denied. Obligation policy check the obligation that define what activities user must do or must not do. A user is obligated to perform some action only when condition is satisfied [16].

### 3.3 Migration Policy

In the migration policy, the user provides all the relevant information needed for the creation role-based schema on the Cloud. The information includes the roles and schemas for the application and the permissions granted to the roles. The user only need to specify their requirements through an interactive interface without taking into consideration of the technical information of the migration data. The XML file is generated from the information provided by the user . This XML file is sent to service providers where database and role based access control information is read and processed[17].

### 3.4 Restriction Policy

The Restraint Policy restricts the number of users per role and number of transactions per role to enhance the security of the architecture so that one id can make a fixed number of transactions, then if the number of transactions will increase the admin will come to know through its monitoring system that unauthorized access has been made and it would be easier to take action against such happening. So the concept of restricting number of transactions per user improves much security and also find the scalability of the system.

### 3.5 Backup and Restoration Policy

In cloud environment backup and restoration policy helps to take the backup of the data in case of any discrepancies. If in any case cloud crashed or does not work properly user will have a backup of the data and can restore it in local server as well as in cloud again. Thus it also helps to make the availability of the data even if the cloud crashes.

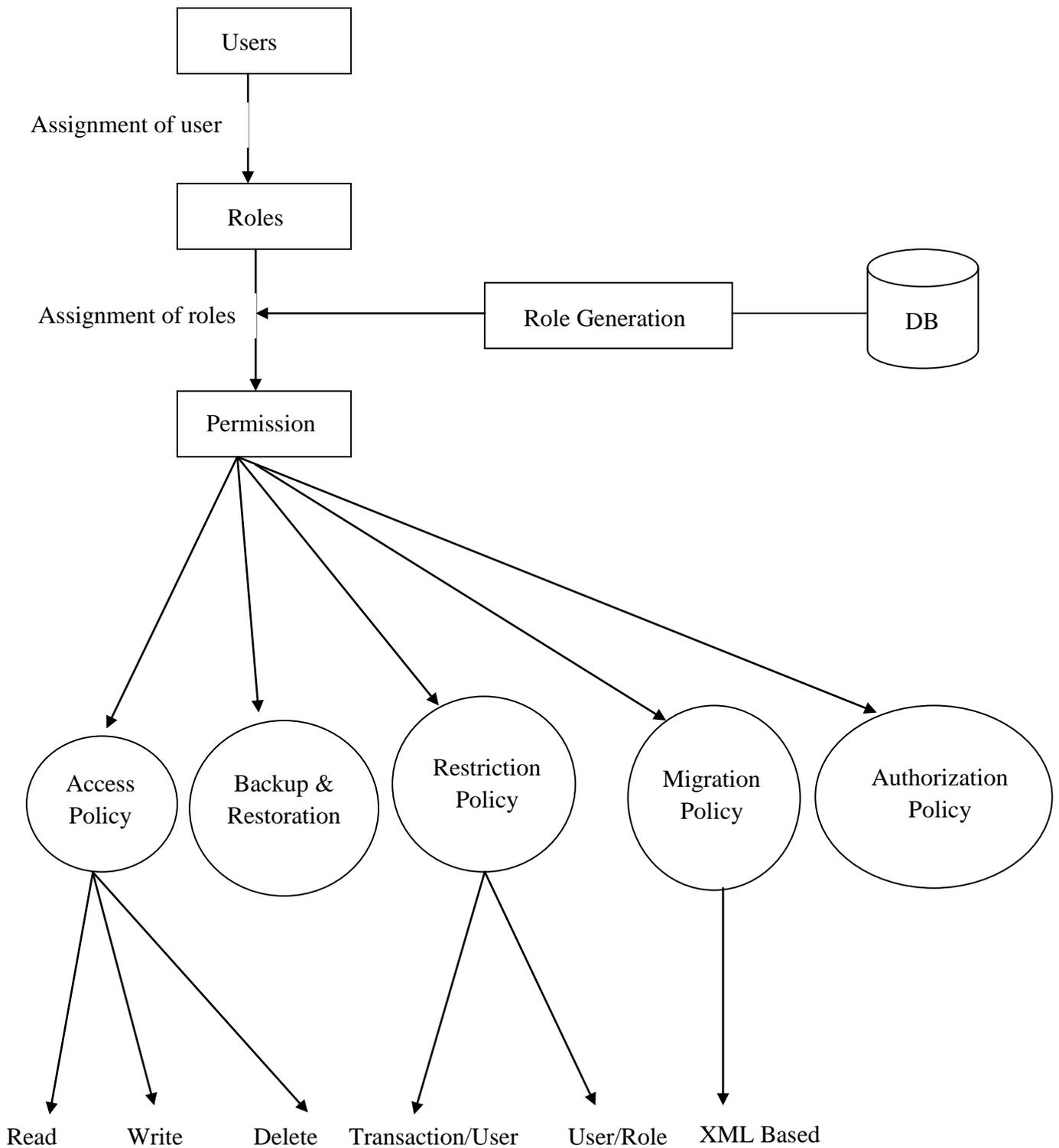

**Figure 2: RBAC with Policy Specification**

## 4. PROPOSED MODEL

Earlier it was not an easy to pick up the entire architecture from one system to another system. Later on when researches were done the developers found their way to migrate the data but the problem was to have secure and efficient migration. Researchers required language which can migrate the data with less complexity and time consumption and now a day's XML is one of the lightest languages . The good thing about XML is that it is supported by every platform. The main objectives of implementing the Role Based access control with policy specification are:

1. Using XML based schema and role migration policy

2. Adding the restriction policy to enhance the security in the cloud

3. To add backup and restoration policy to keep the backup of the existing data.

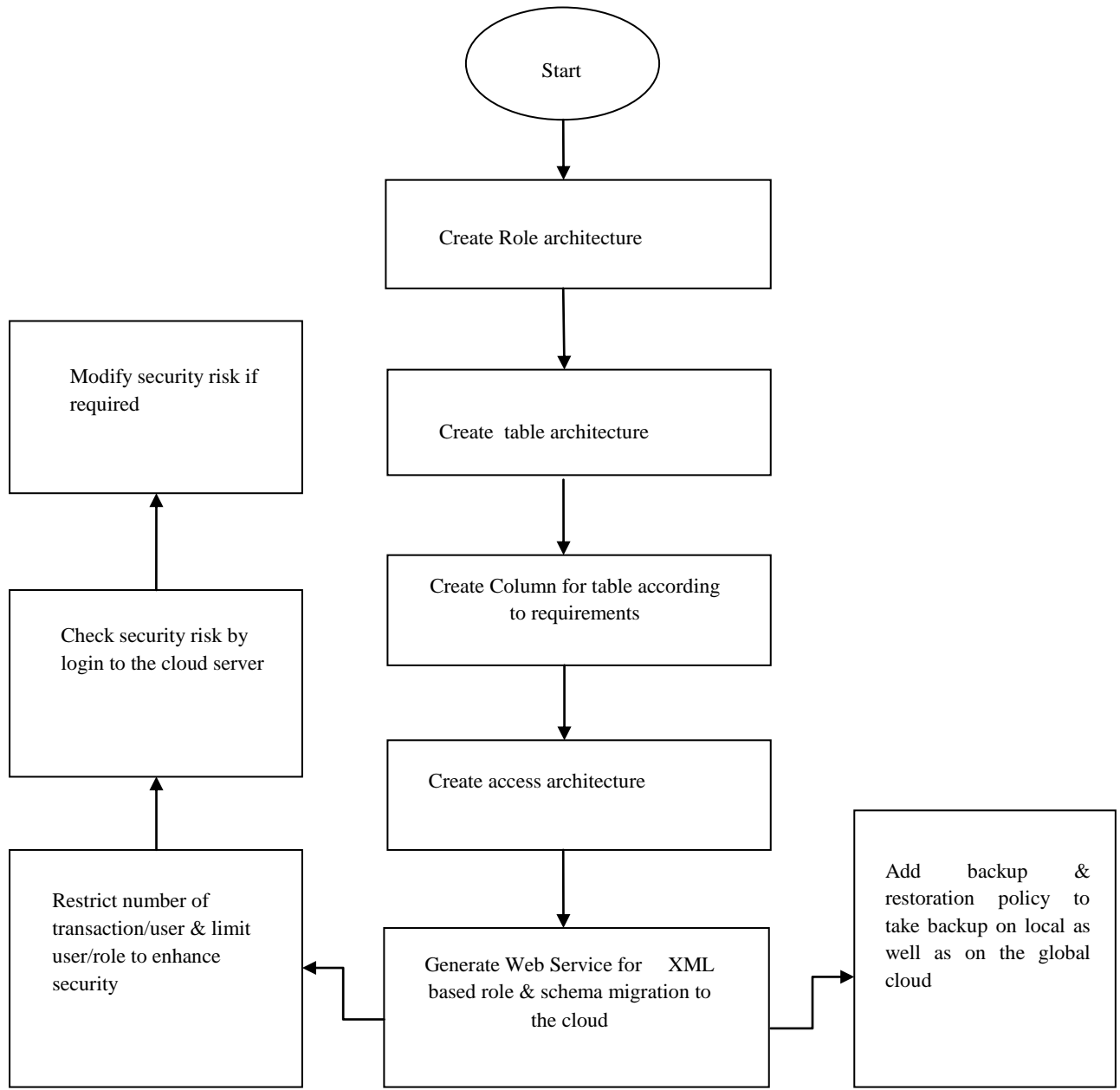

**Figure 3: Work Specification with RBAC on clouds with policy specification**

# 5. CONCLUSION & FUTURE WORK

RBAC with policy specification is proved to be best as compared to other access control model considered in this work because it includes the advantages of traditional RBAC as well as this model has new policies which provide more security and reduce the complexity of system implementation and design. It has been affective in a lot of manners. RBAC architecture saves the data from unauthorized use and modification of the data.

**Table 1 Comparison B/w Normal RBAC v/s RBAC with Policy Specification**

| PARAMETER | NORMAL RBAC | RBAC WITH POLICY |
|---|---|---|
| XML BASED MIGRATION | NO | YES |
| RESTRICTING USER/ ROLE | NO | YES |
| BACKUP & RESTORATION FACILITY | NO | YES |
| TRANSACTION LIMIT | NO | YES |
| SECURITY | LESS | MORE |

This paper proposed various policies for RBAC on clouds for overcoming the limitation of typical access control methods. It is possible to migrate roles and schema using XML. Also it is possible to analyze the role/user ratio according to the position hierarchy, to measure the scalability of the RBAC, with respect to number of roles, number of permissions. RBAC is an architecture which provide the authority to restrict the user if the user is not allowed to go on with the content. There is also concept of backup and restoration of data that helps to make the availability of data even if the main cloud crashes.

While running the application it has been observed that the role based access control model with policy specification is working in the same way as it was working on-premises and roles and database schema has been migrated successfully on the cloud with restriction and backup policy.

As future works privacy, trust and access control based models in the proposed scheme could be considered in RBAC and ontology concept can be added to enhance existing architecture.